\documentclass[twocolumn,showpacs,article]{revtex4}
\usepackage{amsfonts}
\usepackage{amsmath}
\usepackage{amssymb}
\usepackage{graphicx}

\begin{document}

\title{The variety of stable vortical solitons in Ginzburg-Landau media with
radially inhomogeneous losses}
\author{V. Skarka$^{1,2}$, N. B. Aleksi\'{c}$^{2}$, H. Leblond$^{1}$, B. A.
Malomed$^{3}$, and D. Mihalache$^{4}$}
\affiliation{$^{1}$Laboratoire de Photonique d'Angers, EA 4464, Universit\'{e} d'Angers,
2 Boulevard Lavoisier, 49045 Angers Cedex 01, France\\
$^{2}$Institute of Physics, University of Belgrade, 11000 Belgrade, Serbia\\
$^{3}$Department of Physical Electronics, Faculty of Engineering, Tel Aviv
University, Tel Aviv 69978, Israel\\
$^{4}$Horia Hulubei National Institute for Physics and Nuclear Engineering,
407 Atomistilor, Magurele-Bucharest, 077125, Romania}

\begin{abstract}
Using a combination of the variation approximation (VA) and direct
simulations, we consider the light transmission in nonlinearly amplified
bulk media, taking into account the localization of the gain, i.e., the
linear loss shaped as a parabolic function of the transverse radius, with a
minimum at the center. The balance of the transverse diffraction,
self-focusing, gain, and the inhomogeneous loss provide for the hitherto
elusive stabilization of vortex solitons in a large zone of the parameter
space. Adjacent to it, stability domains are found for several novel kinds
of localized vortices, including spinning elliptically shaped ones,
eccentric elliptic vortices which feature double rotation, spinning
crescents, and breathing vortices.
\end{abstract}

\maketitle

\textit{Introduction and the model}. Self-collimated and self-guided light
beams organized as spatial solitons in bulk media are subjects of great
interest \cite{solitons}. In addition to their significance for fundamental
studies, spatial solitons can find applications to the design of all-optical
data-processing schemes. In particular, laser systems may run on dissipative
optical solitons, which are described by the complex Ginzburg-Landau
equations (CGLEs), typically with the cubic-quintic (CQ) nonlinearity,
taking into account the saturable nonlinear gain \cite{CGL}. Crucially
important to the applications is the stability of dissipative solitons.
Families of stable spatial solitons with two transverse dimensions (2D)
exist in materials characterized by a saturable nonlinearity that
compensates the diffraction \cite{Skarka,solitons}. The additional condition
of the balance between losses and gain reduces the family to isolated
solutions, one of which may be stable as an \textit{attractor} \cite%
{attractor,exact}.

The complexity of the CGLEs does not admit exact solutions, with rare
exceptions \cite{exact,delta}. Nevertheless, an analytical approximation,
which provides clues to finding dissipative solitons in the numerical form,
has been developed in the form of the variational approach (VA) adopted for
dissipative systems \cite{VladNajdPRL}. Parameter domains where 2D and 3D
solitons are stable in the CQ-CGLE and related models have been outlined by
means of these methods \cite{VladNajdPRL,Herve}.

Objects of fundamental significance which remained elusive in the studies of
dissipative solitons are solitary vortices. In previous works, such vortices
were found to be stable only in the presence of the diffusion, which occurs
in other physical media, but does not appear in optical systems \cite%
{Lucian,Herve}, nor in dissipative models of Bose-Einstein condensates
(BECs) \cite{Konotop,Kartashov}. The stabilization of multi-peak patterns
carrying vortical phase patterns was also demonstrated by periodic
potentials that may be created in the laser cavity \cite{Herve}.
Nevertheless, the most fundamental circular (``crater-shaped") localized
vortices were not found yet as stable objects in models of optical media. In
this work, we report (for the first time, to our knowledge) \emph{stable}
fundamental vortical modes in a physically relevant model where the
stabilization is provided by an axisymmetric modulation of the linear loss,
with a minimum at the center. In the experiment, the loss modulation may be
induced by a localized gain which partly compensates the uniform loss (an
``iceberg of gain" submerged into the ``sea of loss"). In fact, the gain
applied to a laser cavity is always localized (in particular, due to the
finite width of the pumping beam). Besides the optical media, similar
modulated loss profiles can be engineered in BEC, where they help to support
various matter-wave modes \cite{Konotop}.

We also demonstrate that, when the axisymmetric vortices lose their
stability, they give rise to other novel types of vorticity-carrying modes.
These are spinning elliptically deformed vortices, eccentric elliptic
vortices which feature spinning and precessing, revolving crescents, and
breathing vortices. Thus, the use of the 2D ``iceberg-shaped" gain profile
opens the way to the previously inaccessible class of \emph{stable} vortical
modes in bulk optical media.

We adopt the (2+1)D form of the CQ-CGLE governing the evolution of the
optical field in the bulk medium, $E\left( z,x,y\right) $, along axis $z,$
and the diffraction in the transverse plane \cite{CGL,attractor}:
\begin{gather}
iE_{z}+\frac{1}{2}\left( E_{xx}+E_{yy}\right) +\left( 1-i\varepsilon \right)
|E|^{2}E-\left( \nu -i\mu \right) |E|^{4}E  \notag \\
=-ig(r)E,~r\equiv \sqrt{x^{2}+y^{2}},  \label{CGLE}
\end{gather}%
where positive coefficients $\nu ,\varepsilon $ and $\mu $ account for the
saturation of the Kerr nonlinearity, cubic gain, and quintic loss,
respectively. As said above, the linear-loss coefficient is modulated along
the radial coordinate, $g(r)=\gamma +Vr^{2}$, with $\gamma $, $V>0$. Note
that the limit form of a similarly organized 1D model, with the linear gain
concentrated at a single ``hot spot" (approximated by the delta-function)
and cubic nonlinearity, admits exact stable solutions for pinned solitons
\cite{delta}.

\textit{The variation approximation}. First, we aim to obtain approximate
analytical results, using the VA technique \cite{VladNajdPRL}. To this end,
we adopt the ansatz for vortex soliton with topological charge $1$,
\begin{equation}
E=\frac{A_{\ast }Ar}{R_{\ast }R}\exp \left( -\frac{r^{2}}{2R_{\ast }^{2}R^{2}%
}+i\frac{Cr^{2}}{R_{\ast }^{2}}+i\theta +i\Psi \right) ,  \label{E}
\end{equation}%
where $\theta $ is the angular coordinate, normalizing factors are $A_{\ast
}=3/(2\sqrt{2})$\emph{\ }and\emph{\ }$R_{\ast }=16/9$, while variation
parameters are amplitude $A$, radius $R$, radial chirp $C$, and phase $\Psi $%
. The total power corresponding to ansatz (\ref{E}) is $P\equiv 2\pi
\int_{0}^{\infty }|E(r)|^{2}rdr=$ $P_{\ast }A^{2}R^{2}$, with\emph{\ }$%
P_{\ast }\equiv \pi A_{\ast }^{2}R_{\ast }^{2}$. Skipping straightforward
details, the VA leads to the system of evolution equations:

\begin{equation}
\frac{dA}{dz}=\frac{A}{R_{\ast }^{2}}\left( \frac{5}{2}\varepsilon
A^{2}-2\mu A^{4}-2C\right) -A(\gamma +R_{\ast }^{2}VR^{2}),  \label{A}
\end{equation}%
\begin{equation}
\frac{dR}{dz}=\frac{R}{R_{\ast }^{2}}\left( 2C-\frac{\varepsilon }{2}A^{2}+%
\frac{\mu }{2}A^{4}-R_{\ast }^{4}VR^{2}\right) ,  \label{R}
\end{equation}

\begin{equation}
\frac{dC}{dz}=\frac{1}{R_{\ast }^{2}}\left( \frac{1}{2R^{4}}-\frac{A^{2}}{%
2R^{2}}+\frac{\nu A^{4}}{2R^{2}}-2C^{2}\right) .  \label{C}
\end{equation}%
Steady-state solutions are obtained as fixed points (FPs) of Eqs. (\ref{A})-(%
\ref{C}). In the case of small chirp, it follows from Eq. (\ref{C}) that the
width of the stationary state depends only on its amplitude, $%
R^{2}=(A^{2}-\nu A^{4})^{-1}$, as in conservative systems. In the first
approximation, the small chirp, which makes the dissipative solitons
different from their conservative counterparts, is obtained from Eq. (\ref{R}%
): $C=A^{2}(\varepsilon -\mu A^{2})/4+R_{\ast }^{4}V/\left[ 2A^{2}\left(
1-\nu A^{2}\right) \right] $.

Solving the FP equations following from Eqs. (\ref{A})-(\ref{C})
analytically, we find two different solutions for $A$, as shown in the inset
to Fig. 1. As per general principles of the analysis of dissipative systems,
the solution with larger $A$ may be stable, while its counterpart with
smaller $A$ corresponds to a separatrix between the two attractors \cite%
{exact} -- the stable vortex and trivial state, $E=0$. The solution with
larger $A$ is characterized by $C<0$, which also is a necessary stability
condition for the steady state \cite{VladNajdPRL}. The stability of the FPs
was checked through the computation of eigenvalues for small perturbations,
see the results in Fig. 1.

\begin{figure}[tbp]
\begin{center}
\includegraphics[width=7cm]{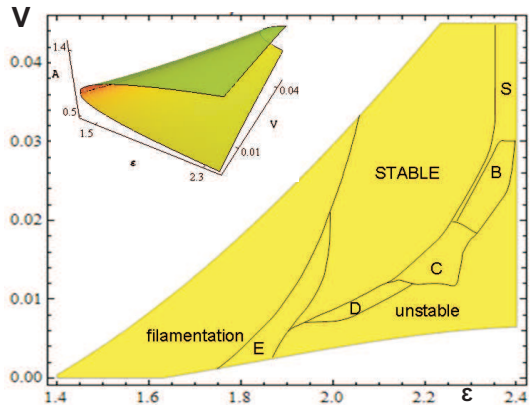}
\end{center}
\caption{(Color online) Stability regions of vortex solitons in plane ($%
\protect\varepsilon ,V$) of the loss-modulation parameters. In this figure
and below, other coefficients in Eq. (\protect\ref{CGLE}) are $\protect%
\gamma =0.29$, $\protect\mu =1.4$, $\protect\nu =0.4$, which adequately
represent the generic situation. The VA predicts the stability in the shaded
(yellow) area, and the corresponding amplitude displayed in the inset.}
\label{Fig1}
\end{figure}

\textit{Numerical results}. The full stability area for the vortex solitons
was identified by means of direct simulations of Eq. (\ref{CGLE}), using the
VA-produced ansatz as the input. The VA predicts the qualitative shape of
the area correctly, although overestimating its size. A typical example of
the fast formation of a stable vortex (which completes by $z=20$, i.e.,
after passing $\lesssim 4$ diffraction lengths) is displayed in Fig. 2.

\begin{figure}[tbp]
\begin{center}
\includegraphics[width=7cm]{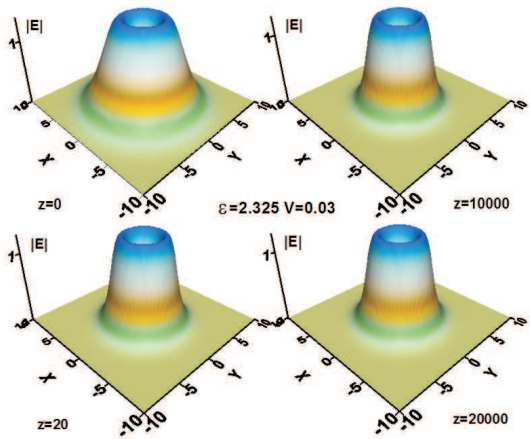}
\end{center}
\caption{(Color online) The self-trapping of a stable vortex from the input
predicted by the VA, at $\protect\varepsilon =2.325$ and $V=0.03$.}
\label{Fig2}
\end{figure}

In the ``filamentation" region in Fig. 1, the modulational instability
breaks the vortex into two segments, at $z\sim 1000$ (closer to the
stability border, this distance increases to $z\sim 3000$). This outcome of
the evolution is explained by the fact that, in this region, the total power
of the vortex ($P$) falls below the breakup threshold, cf. similar
instability scenarios reported in Refs. \cite{Herve}-\cite{Skarka2}.

In region $E$ separating the stability and breakup domains in Fig. 1, $P$
exceeds the breakup threshold, hence the vortex does not split. However, in
this region the evolution leads to squeezing the circular vortex into an
\textit{elliptic} \textit{rotating} one, which remains stable. For instance,
in Fig. 3, the circular vortex persists until $z=3700$. The further
evolution, lasting for $\Delta z\approx 200$, ends up with the transition
into a \emph{stable} elliptic soliton of a larger amplitude, which revolves
with a constant period, $Z\approx 32$. Its profile features a ``volcanic"
shape, with an undulate ``crater" (see the bottom row). This is a novel type
of vorticity-carrying solitons in dissipative media. Note that elliptically
deformed vortices were observed in a conservative nonlocal optical medium
\cite{Moti}, and they were studied theoretically in diverse settings \cite%
{elliptic-theory}, but in those cases the ellipticity was imposed by
anisotropic boundaries, while here the transition to the elliptic shape is
generated \emph{intrinsically}. Rotating solitons of other types (``dipole
propellers") were also predicted in conservative media, but in those cases
the rotation was imparted, rather than spontaneous \cite{propeller}.

\begin{figure}[tbp]
\begin{center}
\includegraphics[width=7cm]{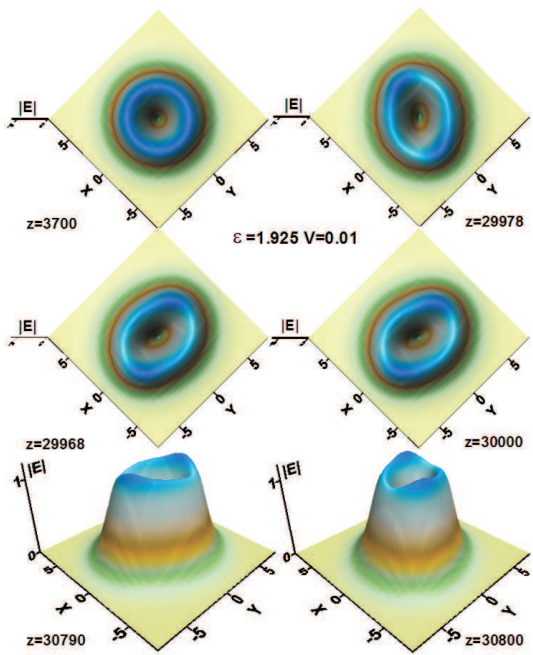}
\end{center}
\caption{(Color online) The self-trapping of a stably rotating elliptic
vortex. The configurations displayed at $z=29968$ and $z=30000$ correspond
to the rotation of the ellipse by $180^{\mathrm{o}}$. Images in the bottom
row display the established ``volcanic" shape of the elliptic vortex.}
\label{Fig3}
\end{figure}

In the right-bottom (``unstable") region of Fig. 1, the vortices suffer
destruction after the propagation distance measured in hundreds of units
(closer to the border of region E, this distance extends to thousands). Four
regions $D,C,B,$ and $S$, which separate\ the stable and unstable regions in
Fig. 1, feature other remarkable vortical patterns. In region $D$, the
stable elliptic vortex spontaneously becomes \textit{eccentric}. As shown in
Fig. 4, it develops the eccentricity after $z=8500$, and after $z=14000$ it
slips out from the central position and starts precessing around it, thus
featuring \emph{double rotation }(the precession and spinning) in the same
direction, with an apparently locked period ratio, $216/54=4$, which may be
caused by a nonlinear parametric resonance in the original spinning elliptic
vortex. The eccentric elliptic vortices remain stable in the course of this
complex motion.

\begin{figure}[tbp]
\begin{center}
\includegraphics[width=7cm]{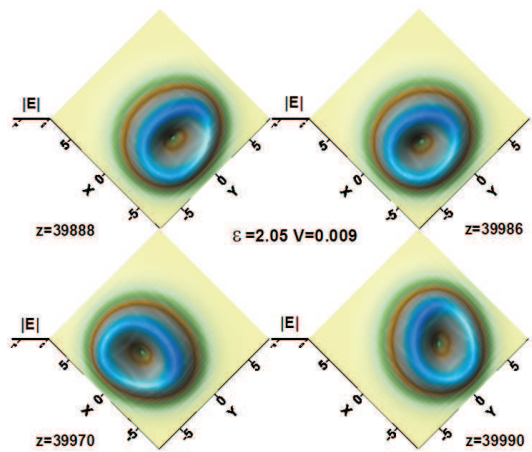}
\end{center}
\caption{(Color online) A stable \textit{eccentric} spinning vortex, which
orbits around the center.}
\label{Fig4}
\end{figure}

Another variety of spontaneously deformed rotating vortices occupies region $%
C$ in Fig. 1. As shown in Fig. 5, after $z=660$ the circular symmetry is
broken by generating a \textit{crescent-shaped} vortex, which fills a half
of the original circle. The crescent rotates around the center, with period $%
Z\approx 12$ (see panels in Fig. 5 for $z=25988$ and $z=26000$). Stable
crescent-shaped solitons were predicted in the conservative model of
rotating BECs trapped in anharmonic axisymmetric potentials \cite{crescent}.

\begin{figure}[tbp]
\begin{center}
\includegraphics[width=7cm]{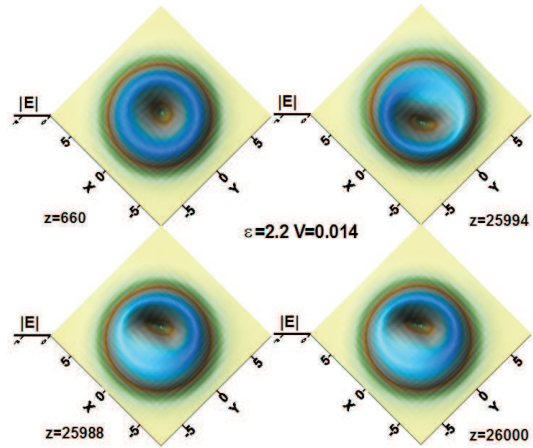}
\end{center}
\caption{(Color online) The evolution of a stable rotating crescent-shaped
vortex.}
\label{Fig5}
\end{figure}

The above species of the vortices feature stable shapes. On the other hand,
in region $B$ of Fig. 1 they are subject to an oscillatory instability,
which immediately transforms them into robust \textit{breathers}, which keep
the vorticity and axial symmetry. Such stable breathers are similar to the
one shown in the first panel of Fig. 6 (the period of its oscillations is $%
Z\approx 5$). Breathing vortices also appear in region $S$; however, as
shown in Fig. 6, they spontaneously develop the elliptic deformation and
cease intrinsic oscillations at $z\simeq 500$. The resulting vortex develops
into an elliptic ``volcano" which rotates with period $Z\approx 26$. The
difference from Fig. 3 is that this vortex species is characterized by the
``crater" which is not undulate but canted.

\begin{figure}[tbp]
\begin{center}
\includegraphics[width=7cm]{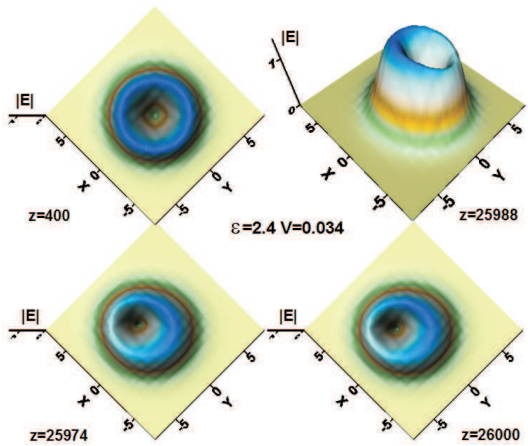}
\end{center}
\caption{(Color online) A vortex breather formed from the VA-predicted input
at $z\simeq 20$ and shown from above at $z=400$, evolves into an
elliptically-shaped ``slanted volcano", whose profile is displayed at $%
z=25988$. It rotates with period $Z\approx 26$ (the images shown at $z=25974$
and $z=26000$ are separated by the period).}
\label{Fig7}
\end{figure}

\textit{Conclusion}. Using the variational approximation and numerical
simulations, we have developed the analysis of the general 2D model of
optical media with the localized linear gain compensated by losses. The
balance of the diffraction, self-focusing, saturable nonlinear gain and the
effective inhomogeneous loss provides for the stabilization of vortex
solitons, which were unstable in previously studied models of isotropic
optical media. In addition to the broad stability area for the vortices,
stability domains are also found for several novel species of vortical
solitons, including elliptic spinning ones, double-rotating eccentric
vortices, revolving crescents, and breathers.

The work suggests extensions in other directions. First, the stabilization
may be also provided by the ``iceberg of gain" elevating above the surface
of the ``sea of loss", i.e., with the gain dominating near the center. In
that case, it is sufficient to have the usual cubic loss, cf. the 1D model
developed in Ref. \cite{delta}. On the other hand, for the stabilization of
vortices it may be relevant to consider the loss-modulation profile with a
minimum at a finite distance from the center. It is also interesting to
consider a possibility of the stabilization of vortices by means of spatial
modulations of the nonlinear loss, which may be relevant to optics and BEC
alike \cite{Kartashov}. Finally, a challenging problem is the stabilization
of 3D vortex solitons.

This work was supported, in a part, by the Ministry of Science of Serbia
under Project No. OI 141031, and by a grant from the High Council for
Scientific and Technological Cooperation between France and Israel.


\begin{thebibliography}{99}
\bibitem{solitons} Y. S. Kivshar and G. P. Agrawal, \textit{Optical
Solitons: From Fibers to Photonic Crystals} (Academic Press, San Diego, CA,
2003); F. Lederer \textit{et al}. 
Phys. Rep. \textbf{436}, 1 (2008); Y. V. Kartashov, V. A. Vysloukh, L.
Torner, in \textit{Progress in Optics} (Elsevier B. V., Amsterdam, 2009),
Vol. 52, p. 63.

\bibitem{CGL}
F. T. Arecchi, S. Boccaletti, and P. L. Ramazza: Phys. Rep. \textbf{318}, 1
(1999); I. S. Aranson and L. Kramer, Rev. Mod. Phys. \textbf{74}, 99 (2002);
N. N. Rosanov,\textit{\ Spatial Hysteresis and Optical Patterns }(Springer,
Berlin, 2002); \textit{Dissipative Solitons}, edited by N. Akhmediev and A.
Ankiewicz, 
(Springer, Berlin, 2005).

\bibitem{Skarka} M. Segev \textit{et al.},
Phys. Rev. Lett. \textbf{68}, 923 (1992); G. S. Duree \textit{et al}.,
\textit{ibid}. \textbf{71}, 533 (1993); M. Segev \textit{et al}.,
\textit{ibid}. \textbf{73}, 3211 (1994); D. N. Christodoulides and M. I.
Carvalho, J. Opt. Soc. Am. B \textbf{12}, 1628 (1995); V. Skarka, V. I.
Berezhiani, and R. Miklaszewski, Phys. Rev. E \textbf{56}, 1080 (1997).

\bibitem{attractor} B. A. Malomed, Physica D \textbf{29}, 155 (1987); S.
Fauve and O. Thual, Phys. Rev. Lett. \textbf{64}, 282 (1990); B. A. Malomed
and H. G. Winful, Phys. Rev. E \textbf{53}, 5365 (1996).

\bibitem{exact} N. N. Akhmediev, V. V. Afanasjev, and J. M. Soto-Crespo,
Phys. Rev. E \textbf{53}, 1190 (1996); J. Atai and B. A. Malomed, Phys.
Lett. A \textbf{246}, 412 (1998); W. J. Firth and P. V. Paulau, Eur. Phys.
J. D \textbf{59}, 13 (2010).

\bibitem{delta} C.-K. Lam \textit{et al.},
Eur. Phys. J. Special Topics \textbf{173}, 233 (2009); C. H. Tsang \textit{%
et al.}, 
Eur. Phys. J. D \textbf{59}, 81 (2010).

\bibitem{VladNajdPRL} V. Skarka, and N. B. Aleksi\'{c}, Phys. Rev. Lett.
\textbf{96}, 013903 (2006).

\bibitem{Herve} D. Mihalache \textit{et al}.,
Phys. Rev. A \textbf{75}, 033811 (2007); \textit{ibid}. \textbf{77}, 033817
(2008); \textbf{81}, 025801 (2010); Phys. Rev. E \textbf{78}, 056601 (2008);
H. Leblond, B. A. Malomed, and D. Mihalache,
Phys. Rev. A \textbf{80}, 033835 (2009).

\bibitem{Konotop} V. A. Brazhnyi \textit{et al}., 
Phys. Rev. Lett. \textbf{102}, 144101 (2009); Y. V. Bludov and V. V.
Konotop, Phys. Rev. A \textbf{81}, 013625 (2010).

\bibitem{Kartashov} F. Kh. Abdullaev \textit{et al}.,
Phys. Rev. A \textbf{76}, 043611 (2007); Y. V. Kartashov \textit{et al}.,
Opt. Lett. \textbf{35}, 1638 (2010).

\bibitem{Lucian} L.-C. Crasovan, B. A. Malomed, and D. Mihalache,
Phys. Rev. E \textbf{63}, 016605 (2000).

\bibitem{Skarka2} V. Skarka \textit{et al.}, 
Phys. Rev. B \textbf{81}, 035202 (2010).

\bibitem{Moti} C. Rotschild 
\textit{et al}., Phys. Rev. Lett. \textbf{95}, 213904 (2005).

\bibitem{elliptic-theory} J. J. Garc\'{\i}a-Ripoll \textit{et al}.,
Phys. Rev. Lett. \textbf{87}, 140403 (2001); F. Ye, D. Mihalache, and B. Hu,
Phys. Rev. A \textbf{79}, 053852 (2009); F. Ye
\textit{et al}., J. Opt. Soc. Am. B \textbf{27}, 757 (2010).

\bibitem{propeller} T. Carmon \textit{et al}.,%
Phys. Rev. Lett. \textbf{87}, 143901 (2001); B. A. Malomed \textit{et al.},
Phys. Rev. A \textbf{\ 70}, 043616 (2004).

\bibitem{crescent} E. Lundh, A. Collin, and K.-A. Suominen, Phys. Rev. Lett.
\textbf{92}, 070401 (2004); G. M. Kavoulakis, A. D. Jackson, and G. Baym,
Phys. Rev. A \textbf{70}, 043603 (2004); A. Collin, \textit{ibid}. \textbf{73%
}, 013611 (2006); H. Sakaguchi and B. A. Malomed, \textit{ibid}. \textbf{78}%
, 063606 (2008).
\end{thebibliography}
\end{document}